\UseRawInputEncoding
\documentclass[aps,showpacs,prep,graphics]{revtex4}

\usepackage{amssymb}
\usepackage[dvips]{graphicx}
\usepackage{amsmath}
\usepackage{bm}
\usepackage{epsfig}
\usepackage{graphicx}
\usepackage{caption}
\usepackage{wrapfig}
\usepackage{color}

\begin{document}

\title{Dark energy from a geometrical gauge scalar-tensor theory of gravity}

\author{ $^{1}$ José Edgar Madriz Aguilar, $^{1}$ M. Montes, $^{2}$ A. Bernal
\thanks{E-mail address: mariana.montnav@gmail.com} }
\affiliation{$^{1}$ Departamento de Matem\'aticas, Centro Universitario de Ciencias Exactas e ingenier\'{i}as (CUCEI),
Universidad de Guadalajara (UdG), Av. Revoluci\'on 1500 S.R. 44430, Guadalajara, Jalisco, M\'exico,  \\
and\\
$^{2}$ Departamento de F\'isica, Centro Universitario de Ciencias Exactas e ingenier\'{i}as (CUCEI),
Universidad de Guadalajara (UdG), Av. Revoluci\'on 1500 S.R. 44430, Guadalajara, Jalisco, M\'exico. \\
E-mail:  
 mariana.montes@academicos.udg.mx, alfonso.bernal@alumnos.udg.mx}

\begin{abstract}

 In this paper we obtain some cosmological solutions that describe the present period of accelerating expansion of the universe in the framework of a geometrical gauge scalar-tensor theory of gravity. The background geometry in the model is the Weyl integrable and we found a class of power law solutions for the Weyl scalar field when an invariant metric is employed in a power law expanding universe. We obtain a deceleration  and an equation of state parameters in agreement with PLANCK 2018 observational data for some specific parameters of the model. The deceleration parameter tends asymptotically to $-1/2$. 
\end{abstract}

\pacs{04.50. Kd, 04.20.Jb, 02.40k, 98.80.−k, 98.80.Jk}
\maketitle

\vskip .5cm
 Weyl-Integrable geometry, accelerating expansion of the universe, gauge scalar-tensor gravity.

\section{INTRODUCTION}

Since 1998 it was discovered that in the present epoch the universe is expanding in an accelerated manner \cite{c1,c2,c3}. The origin of such acceleration has become one of the greatest challenges of modern cosmology. A positive cosmological constant has been one of the first options that appeared to explain such acceleration at least  in general relativity \cite{c4}. However, a cosmological constant can be viewed as the existence of a vacuum energy , called dark energy, responsible for the acceleration \cite{c5}. When this vacuum energy is described by the quantum field theory then appears the well-known cosmological constant problem \cite{c6,c7}. With the intention to avoid this problem many other proposals to explain the acceleration have appeared.  Dynamical cosmological constant \cite{c8,c9,c10,c11} and different versions of quintessence models \cite{c12,c13,c14} can be found within these proposals . An interesting feature of the class of quintessence mo\-dels is that dark energy is driven by a scalar field similar to what happens in the early inflationary epoch of the universe. However one important question around this is about the origin of the scalar field. Inspired in this issue some cosmological models in which the scalar field employed is part of the affine structure of the geometry have recently appeared. A particular class of these models is based on the so called geometrical scalar-tensor theories of gravity. \\

Geometrical scalar-tensor theories of gravity have been first proposed by C. Romero and collaborators \cite{c15,c16}. They started considering a typical scalar-tensor theory of gravity and instead of imposing the background geometry they adopted the Palatini's variational principle to determine such geometry. 
The result is that the natural backgroud geometry of this class of theories is determined by the Weyl-integrable compatibility condition. An extension of this idea was introduced by J. E. Madriz-Aguilar and collaborators by considering that the action of the theory must have the same symmetry group than the background geometry. To achieve it they introduce a gauge covariant derivative and  construct a new invariant action. This theory is also known as geometrical gauge scalar-tensor theory of gravity \cite{c9,c13}. Different topics have studied in the light of these frameworks, for example interacting quintessence scenarios \cite{c13}, scalar fluctuations of the metric during inflation \cite{c14}, generation of the seeds of cosmic magnetic fields during inflation \cite{c15}, Higgs inflation frameworks \cite{c14,c16}, $(1+2)$-dimensional gravity \cite{c17} and viscous cosmology \cite{c18}, among others. \\

 The idea of considering scalar fields geometrical in origin have also been employed in other models. For example in some approaches in which the Palatini's principle is employed \cite{c19, c20}, and in the so named conformal equivalence principle \cite{c21}. Geometrical scalar fields have been even used in some theories of gravity with extra dimensions. Induced matter theory and relativistic quantum approaches in non-Riemannian geometries are good examples \cite{c22, c23}. The interest in non-riemannian geometries and in particular in the Weyl class of geometries has recently increased due to their proposal to solve for example the frame equivalence issue. A disformal generalization of a Weyl structure is studied in \cite{c24}. Some cosmological models using scalar fields in Weyl geometry can be also  in \cite{c25}.\\

In this paper we obtain cosmological solutions in the framework of a geometrical gauge scalar-tensor theory of gravity capable to describe the present period of accelerating expansion in the universe. The paper is organized as follows. Section I is left for the introduction. In section II we derive the field equations of the geometrical gauge scalar-tensor theory of gravity. In section III we reduce the field equations to the case of cosmological scales. Section IV is devoted to obtain solutions of the cosmological field equations that describe the present period of accelerating expansion. Finally in section V we give some final comments as conclusions.\\

\section{The field equations}

Let us start writing the general  action of a  scalar-tensor theory of gravity  in the form \cite{c13}
\begin{equation}\label{anne1}
    {\cal S}=\int\,d^{4}x\,\sqrt{-h}\left[e^{-\varphi}\left( \frac{R}{16\pi G}-\frac{1}{2}\omega(\varphi)h^{\alpha\beta}\varphi_{,\alpha}\varphi_{,\beta}\right)+V(\varphi)\right],
\end{equation}
with $\omega(\varphi)$ being a well behaved function of the scalar field $\varphi$, $V(\varphi)$ is the scalar potential,  $h_{\alpha\beta}$ is the metric tensor, $g=det(g_{\alpha\beta})$ and $R(\Gamma)$ is the Ricci scalar of curvature that we consider depending on the affine connection. Palatini variational principle leads to the compatibility condition  
\begin{equation}\label{ane1}
    \nabla_{\mu} h_{\alpha\beta}=\varphi_{,\mu}h_{\alpha\beta},
\end{equation}
which corresponds to a Weyl-Integrable background geometry and $\nabla_{\mu}$ is denoting the covariant derivative with respect the  Weyl connection coordinate components
\begin{equation}\label{ane2}
    ^{(w)}\Gamma^{\alpha}_{\mu\nu}=\,\!^{(R)}\Gamma^{\alpha}_{\mu\nu}-\frac{1}{2}\left(\varphi_{,\nu}\delta^{\alpha}_{\mu}+\varphi_{,\mu}\delta^{\alpha}_{\nu}-\varphi_{,\sigma}h^{\alpha\sigma}h_{\mu\nu}\right),
\end{equation}
where $^{(R)}\Gamma^{\alpha}_{\mu\nu}$ is de Levi-Civita connection. The expression \eqref{ane1} is invariant under the transformations
\begin{eqnarray}\label{ane3}
\bar{h}_{\alpha\beta} &=&  e^{f}h_{\alpha\beta},\\
\label{ane4}
\bar{\varphi} &=& \varphi + f,
\end{eqnarray}
with $f(x^{\alpha})$ is a well-behaved function of the space-time coordinates and $k$ is a constant. Notice that under \eqref{ane3} and \eqref{ane4} applied simultaneously the Ricci tensor remains invariant
\begin{equation}
    \bar{R}_{\mu\nu}(\bar{h}_{\alpha\beta},\bar{\varphi})=R_{\mu\nu}(h_{\alpha\beta},\varphi),
\end{equation}
whereas the scalar curvature changes according to the rule
\begin{equation}
    \bar{R}(\bar{h},\bar{\varphi})=e^{-f}\,R(h,\varphi).
\end{equation}

However, it is not difficult to verify that the action \eqref{anne1} does not remain invariant under \eqref{ane3}-\eqref{ane4}. Hence, an  invariant  action results to be
\begin{eqnarray}\label{Camp1}
{\cal S}=\int d^4x\,\sqrt{-h}\,\,e^{-\varphi}\,\left[\frac{R}{16\pi G}-\frac{1}{2}\omega(\varphi)h^{\alpha\beta}\varphi_{:\alpha}\varphi_{:\beta}+V(\varphi)e^{-\varphi}-\frac{1}{4}e^{\varphi}h^{\mu\alpha}h^{\nu\beta}H_{\mu\nu}H_{\alpha\beta}\right],\label{AE1}
\end{eqnarray}
where we have introduced the gauge weylian covariant derivative 
\begin{equation}\label{ane5}
    \varphi_{:\mu}=\nabla_\mu\varphi+\gamma B_{\mu}\varphi,
\end{equation}
with $\gamma$ being an imaginary constant, $B_{\lambda}$ is a gauge field defined in every point of the space-time manifold and $H_{\alpha\beta}=\partial_{\alpha}(\varphi B_{\beta})-\partial_{\beta}(\varphi B_{\alpha})$ is a strength field. The action \eqref{AE1} is invariant only when the next transformations are valid
\begin{eqnarray}\label{ane6}
\bar{\varphi}\bar{B}_{\mu} &=&\varphi B_{\mu}-\gamma^{-1}f_{,\mu}\\
\label{ane7}
\bar{V}(\bar{\varphi}) &=& V(\bar{\varphi}-f)=V(\varphi),\\
\label{ane8}
\bar{\omega}(\bar{\varphi}) &=& \omega(\bar{\varphi}-f) = \omega(\varphi).
\end{eqnarray}
The expression \eqref{ane6} corresponds to the transformation of the elements of the algebra of the  group $U(1)$. Thus, we can associate the field $W_{\mu}=\varphi B_{\mu}$ with an electromagnetic potential defined on the Weyl-integrable background geometry. In this manner the action \eqref{Camp1} is invariant under the set of transformations \eqref{ane3}, \eqref{ane4} and 
\begin{equation}\label{transg}
\overline W_\mu=W_\mu-\gamma^{-1}f_{,\mu}.
\end{equation}
The field equations derived from the action \eqref{AE1} read
\begin{eqnarray}
&&  e^{-\varphi}\,^{(w)}G_{\alpha\beta}+2e^{-\varphi}\left(\nabla_{\alpha}\nabla_{\beta}\varphi-2\nabla_{\alpha}\varphi\nabla_{\beta}\varphi+2h_{\alpha\beta}\nabla_{\sigma}\varphi\nabla^{\sigma}\varphi-h_{\alpha\beta}\,^{(w)}\Box\varphi\right)=8\pi G e^{-\varphi}\left[\omega(\varphi)\varphi_{:\alpha}\varphi_{:\beta}\right.\nonumber \\
&& \left. -
\frac{1}{2}h_{\alpha\beta}\left(\omega\varphi^{:\sigma}\varphi_{:\sigma}-2V(\varphi)\,e^{-\varphi}\right)\right]+8\pi G\left(h^{\nu\mu}H_{\alpha\nu}H_{\beta\mu}-\frac{1}{4}\, h_{\alpha\beta} H^{\mu\nu}H_{\mu\nu}\right), \label{AE5}\\
&&\left(\omega^{\prime}(\varphi)-\omega(\varphi)\right)\nabla^{\alpha}\varphi\nabla_{\alpha}\varphi+\omega(\varphi)\Box\varphi+\gamma\left(\omega^{\prime}(\varphi)-\omega(\varphi)\right)\varphi\nabla^{\alpha}\varphi A_{\alpha}+\gamma\varphi\,\,\omega(\varphi)\,{\cal D}_{\alpha}A^{\alpha}-\frac{^{(w)}\!R}{16\pi G}≈\nonumber\\
&&+\frac{1}{2}\left(\omega(\varphi)-\omega^{\prime}(\varphi)\right)\varphi^{:\alpha}\varphi_{:\alpha}+e^{-\varphi}(V^{\prime}(\varphi)-2V(\varphi))=0,\label{AE6}\\
&& {\cal D}_\alpha\,H^{\alpha\beta}=\gamma e^{-\varphi}\omega(\varphi)\varphi^{:\beta},\label{AE7}
\end{eqnarray}
where $^{(w)}G_{\alpha\beta}$ is de Weylian Einstein tensor, $^{(w)}R$ is the Weylian scalar curvature,  $\nabla_{\alpha}$ is denoting the Weylian covariant derivative, $D_{\alpha}$ stands for the Riemannian covariant derivative,  $^{(w)}\Box$ is the Weylian D'Alambertian operator, $\Box$ denotes the Riemannian D'Alambertian operator and the prime is denoting derivative with respect $\varphi$. Taking the trace of \eqref{AE5}, the equation \eqref{AE6} written in terms of the field $W^{\mu}$ reads
\begin{eqnarray}
 && \left(\omega^{\prime}(\varphi)-\omega(\varphi)-\frac{5}{8\pi G}\right)\nabla_{\alpha}\varphi\nabla^{\alpha}\varphi+\left(\omega(\varphi)-\frac{3}{8\pi G}\right)\Box\varphi +\gamma\left(\omega^{\prime}(\varphi)-\omega(\varphi)\right)\nabla_{\alpha}\varphi \,W^{\alpha}+\gamma\omega(\varphi)\left({\cal D}_{\mu}W^{\mu}-\nabla_{\mu}(\ln\varphi)\,W^{\mu}\right)\nonumber\\
  \label{qrc1}
 && -\left(\omega(\varphi)-\frac{1}{2}\omega^{\prime}(\varphi)\right)\varphi^{:\alpha}\varphi_{:\alpha}+e^{-\varphi}V^{\prime}(\varphi)=0.
\end{eqnarray}
As we have mentioned the background geometry of these equations is of  the Weyl-integrable type and in this setting  both $\varphi$ and $W_{\mu}$ can be considered   geometrical fields that form part of the affine structure of the space-time. In the next section we shall obtain  cosmological applications of the system \eqref{AE5}-\eqref{AE7}, in particular we will focus on those that describe the current acceleration in the expansion of the universe.


\section{The cosmological field equations}

In order to study some cosmological implications of the field equations \eqref{AE5}-\eqref{AE7}, we consider that the Weyl scalar field can be expressed by the sum of two contributions, one on cosmological scales that respects the cosmological principle and another one valid on smaller scales, thus we assume the separation formula
\begin{equation}
    \label{qrc2}
    \varphi(t,x^{i})=\phi(t)+\zeta(t,x^{i}),\qquad |\zeta|\ll |\phi|,
\end{equation}
where $\phi(t)$ is the cosmological part of the field and $\zeta(t,x^{i})$ accounts for the contribution of the Weyl field on non-cosmological scales. The gauge field $W^{\mu}$ is a vector field, thus in order to maintain valid the cosmological principle we will consider that its contribution is only on non-cosmological scales. Hence, it follows from \eqref{AE5}-\eqref{AE7} and \eqref{qrc2} that the field equations valid on cosmological scales acquire the form
\begin{eqnarray}
&& ^{(w)}\!G_{\alpha\beta}-4\left(\nabla_{\alpha}\phi\nabla_{\beta}\phi-h_{\alpha\beta}\nabla_{\sigma}\phi\nabla^{\sigma}\phi\right)+2\nabla_{\alpha}\nabla_{\beta}\phi-2h_{\alpha\beta}^{(w)}\Box\phi=8\pi G [\omega(\phi)\nabla_{\alpha}\phi\nabla_{\beta}\phi-\frac{h_{\alpha\beta}}{2}\left(\omega(\phi)\nabla^{\sigma}\phi\nabla_{\sigma}\phi-2e^{-\phi}V(\phi)\right)],\nonumber\\
\label{mgc1}\\
 && \left(\frac{3}{2}\omega^{\prime}(\phi)-2\omega(\phi)-\frac{5}{8\pi G}\right)\nabla^{\alpha}\phi\nabla_{\alpha}\phi+\left(\omega(\phi)-\frac{3}{8\pi G}\right)\Box\phi+e^{-\phi}V^{\prime}(\phi)=0,
 \label{mgc2}
\end{eqnarray}
where here the prime is denoting derivative with respect to $\phi$.\\

Now, in order to introduce matter sources in our cosmological model it is important to remember that in this geometrical  formalism gravity is described at the same time by the metric $h_{\alpha\beta}$ and the scalar field $\varphi$ \cite{c13}. Thus, it follows that under the consideration that gravity couples to matter then both fields do. This fact is expressed in the matter action
\begin{equation}
    \label{qna1}
S_m=\int \sqrt{-h}\,e^{-2\varphi}\,L_{m}(e^{-\varphi}h_{\alpha\beta},\chi,\,^{(w)}\nabla\chi),
\end{equation}
where $\chi$ denotes a matter field and $L_m$ is a matter lagrangian constructed in the same way that it is done in special relativity. In this manner the energy momentum tensor $T_{\mu\nu}(\varphi,h_{\alpha\beta},\chi,\,^{(w)}\nabla\chi)$ is determined by means of the formula
\begin{equation}
\label{qna2}
    \delta\int d^{4}x\sqrt{-h}\,e^{-2\phi}L_{m}(e^{-\varphi}h_{\alpha\beta},\chi,\,^{(w)}\nabla\chi)=\int d^{4}x\sqrt{-h}\,e^{-2\varphi}\, T_{\mu\nu}(\varphi,h_{\alpha\beta},\chi,\,^{(w)}\nabla\chi)\delta(e^{\varphi}h^{\mu\nu}),
\end{equation}
where $\delta$ is denoting variational with respect to $h_{\mu\nu}$ and $\varphi$.\\ 

On the other hand, it is not difficult to see that due to the Weyl transformations the differential line element defined with the metric $h_{\alpha\beta}$ is not an invariant. Thus, we introduce  the invariant metric $g_{\alpha\beta}=e^{-\varphi}h_{\alpha\beta}$ which on cosmological scales reduces to $g_{\mu\nu}=e^{-\phi}h_{\alpha\beta}$. Hence, in terms of the invariant metric and in presence of matter sources the field equations \eqref{mgc1} and \eqref{mgc2} read  
\begin{eqnarray}
 \label{qna3}
 G_{\alpha\beta}-4({\cal D}_{\alpha}\phi{\cal D}_{\beta}\phi-g_{\alpha\beta}{\cal D}_{\sigma}\phi{\cal D^{\sigma}\phi})+2{\cal D}_{\alpha}{\cal D}_{\beta}\phi -2g_{\alpha\beta}\,\Box\phi &=& 8\pi G T^{(m)}_{\alpha\beta}+8\pi G T^{(\phi)}_{\alpha\beta},\\
 \label{qna4}
 \omega(\phi)\,\Box\phi+\frac{1}{2}\omega^{\prime}(\phi)g^{\mu\nu}{\cal D}_{\mu}\phi{\cal D}_{\nu}\phi+V^{\prime}(\phi) &=& 0,
\end{eqnarray}
where here the prime is denoting derivative with respect to $\phi$, the energy-momentum tensor for matter sources is denoted by $T_{\alpha\beta}^{(m)}$ and 
\begin{equation}
    \label{qna5}
    T^{(\phi)}_{\alpha\beta}=\omega(\phi){\cal D}_{\alpha}\phi {\cal D}_{\beta}\phi -\frac{1}{2}g_{\alpha\beta}\left(\omega(\phi){\cal D}_{\sigma}\phi{\cal D}^{\sigma}\phi-2e^{-\phi}V(\phi)\right),
\end{equation}
stands for the energy-momentum tensor associated with the scalar field $\phi$. \\

Now, we consider the Friedmann-Lemaitre-Robertson-Walker (FLRW) line element for a spatially-flat universe as
\begin{eqnarray}\label{AE11}
ds^2=dt^2-a^2(t)(dx^2+dy^2+dz^2)
\end{eqnarray}
being $a(t)$ the cosmological scale factor and $t$ the cosmic time. As usually we assume that the matter content of the universe can be described by an energy-momentum tensor corresponding to a perfect fluid 
\begin{equation}\label{qna6}
    T^{(m)}\,^{\alpha}\,_{\beta}=diag(\rho_T,-p_T,-p_T,-p_T)
\end{equation}
 where $\rho_T=\rho_m+\rho_r$ and $p_T=p_r$ stand for the total energy density and the total pressure of the cosmic fluid, respectively. The energy density of matter and radiation are denoted respectively by $\rho_m$ and $\rho_r$, whereas $p_r$ accounts for the radiation pressure.  Thus for a comovil class of observers $U^{\lambda}=\delta^{\lambda}_0$
related to the metric background  \eqref{AE11} the  equations \eqref{qna3} yield 
\begin{eqnarray}
&&3H^2=8\pi G (\rho_m+\rho_r)+8\pi G \left(\rho_{\phi}+\frac{6H\dot{\phi}}{8\pi G}\right)\label{AE12}\\
&& 2\frac{\ddot{a}}{a}+H^2=-8\pi G p_r-8\pi G\left(p_{\phi}-\frac{2\ddot{\phi}+4H\dot{\phi}-4\dot{\phi}^2}{8\pi G}\right)
\label{AE13}
\end{eqnarray}
where $H=\dot a/a$ is the Hubble parameter, the dot denotes derivative with respect the cosmic time t and 
\begin{eqnarray}
 && \rho_{\phi}=\frac{1}{2}\omega(\phi)\dot{\phi}^2+V(\phi),\label{qna7}\\
 && p_{\phi}=\frac{1}{2}\omega(\phi)\dot{\phi}^2-V(\phi),\label{qna8}
\end{eqnarray}
are the energy density and pressure associated to the scalar field $\phi$. 
Similarly, \eqref{qna4} now reads
\begin{equation}\label{qna9}
    \omega(\phi)(\ddot{\phi}+3H\dot{\phi})+\frac{1}{2}\omega^{\prime}(\phi)\dot{\phi}^2+V^{\prime}(\phi)=0.
\end{equation}
It follows from \eqref{AE12} and \eqref{AE13} that the dark energy sector is governed by the $\rho_{\phi}$, $p_{\phi}$ and extra terms. Thus we can introduce the quantities
\begin{eqnarray}
 \rho_{de} &=&  \rho_{\phi}+\frac{6H\dot{\phi}}{8\pi G},\label{qna10}\\
 p_{de} &=& p_{\phi}-\frac{2\ddot{\phi}+4H\dot{\phi}-4\dot{\phi}^2}{8\pi G},\label{qna11}
\end{eqnarray}
describing the energy density and pressure for the dark energy sector in the model.

\section{Cosmological solutions exhibiting accelerated expansion}

Now, let us to obtain the conditions for accelerated
expansion solutions of the model. Thus, it follows from \eqref{AE12} and \eqref{AE13} the acceleration equation
\begin{equation}
    \label{qcu1}
    \frac{\ddot{a}}{a}=-4\pi G\left[p_r+\frac{1}{3}(\rho_m+\rho_r)+p_{\phi}+\frac{1}{3}\rho_{\phi}-\frac{2\ddot{\phi}-H\dot{\phi}+4\dot{\phi}^2}{8\pi G}\right].
\end{equation}
It can be easily seen from \eqref{qcu1} that to achieve
$\ddot a>0$ solutions, must be valid the condition
\begin{equation}\label{e18}
\frac{2}{3}\Omega_r+\frac{1}{3}\Omega_m+\left(\omega_{de}+\frac{1}{3}\right)\Omega_{de}<0,
\end{equation}
where $\Omega_r$ $\Omega_m$ and $\Omega_{de}$ are the radiation, matter and dark energy density  parameters, respectively. The equation of state parameter for the dark energy sector is given by
\begin{equation}
    \label{qcu2}
    \omega_{de}=\frac{8\pi G\left(\frac{1}{2}\omega(\phi)\dot{\phi}^{2}-V(\phi)\right)-2\ddot{\phi}-4H\dot{\phi}+4\dot{\phi}^2}{8\pi G\left(\frac{1}{2}\omega(\phi)\dot{\phi}^2+V(\phi)\right)+6H\dot{\phi}}.
\end{equation}
Given that $2/3\,\Omega_{r}+1/3\,\Omega_m >0$, the inequality \eqref{e18} is satisfied when
\begin{eqnarray}
 \label{qcu3}
 \omega_{de}+\frac{1}{3} &<& 0,\\
 \label{qcu4}
 \left|\left(\omega_{de}+\frac{1}{3}\right)\Omega_{de}\right| &>& \frac{2}{3}\Omega_{r}+\frac{1}{3}\Omega_{m}.
\end{eqnarray}
The condition \eqref{qcu3} can be put in the form
\begin{equation}
\label{qcu5}
   \frac{8\pi G\left(\frac{1}{2}\omega(\phi)\dot{\phi}^{2}-V(\phi)\right)-2\ddot{\phi}-4H\dot{\phi}+4\dot{\phi}^2}{8\pi G\left(\frac{1}{2}\omega(\phi)\dot{\phi}^2+V(\phi)\right)+6H\dot{\phi}}<-\frac{1}{3}. 
\end{equation}
On the other hand, by means of the auxiliary field
\begin{equation}\label{qcu6}
    \Phi=\int \sqrt{\omega(\phi)}\,d\phi,
\end{equation}
the equation \eqref{qna9} becomes
\begin{equation}\label{qcu7}
    \ddot{\Phi}+3H\dot{\Phi}+U^{\prime}(\Phi)=0,
\end{equation}
where $U(\Phi)=V(\Phi)$ is the potential in terms of the new field $\Phi$ and the prime here is denoting derivative with respect to $\Phi$. Now, in order to illustrate the formalism we consider a power law expansion of the universe. Thus the Hubble parameter and the scale factor read respectively as
\begin{equation}
    \label{qcu8}
    H(t)=\frac{p}{t},\qquad a(t)=a_0\left(\frac{t}{t_0}\right)^p,
\end{equation}
being $p>1$ a parameter characterizing the expansion, $t_0$ is the present time and $a_0=a(t_0)$. In view of \eqref{qcu8} and as it is usually done in scalar-tensor theories we consider a power-law form for the scalar field $\Phi$ as follows
\begin{equation}\label{qcu9}
    \Phi(t)=\Phi_0\left(\frac{t_0}{t}\right)^{n},
\end{equation}
where $\Phi_0=\Phi(t_0)$ and $n$ is an integer parameter. For the choices made to satisfy the equation \eqref{qcu7}, inserting \eqref{qcu8} and \eqref{qcu9} in \eqref{qcu7} we obtain the potential
\begin{equation}
    \label{qcu10}
    U(\Phi)=U_0\left(\frac{\Phi}{\Phi_0}\right)^{\frac{2(1+n)}{n}},
\end{equation}
being $U_0=U(\Phi_0)$. Now, the simplest case we can study is when  $\omega(\phi)=\omega_0=const.$ In this particular example the expressions \eqref{qcu6} and \eqref{qcu10} lead to the potential
\begin{equation}\label{qcu11}
    V(\phi)=U_0\left(\frac{\phi}{\phi_0}\right)^{\frac{2(1+n)}{n}}.
\end{equation}
With the help of \eqref{AE12}, \eqref{AE13}, \eqref{qcu8}, \eqref{qcu9} and \eqref{qcu10} the deceleration parameter results
\begin{equation}
    \label{qcu12}
    q=-\left[1-\frac{1}{2}\frac{\Omega_m+\frac{4}{3}\Omega_r+\left(1+\frac{1}{2\pi G\omega_0}\right)\left(\frac{8\pi G}{3H^2}\right)n^2\Phi_0^2t_0^{2n}t^{-2(1+n)}-(np+n(n+1))\frac{2\Phi_0t_0^n}{3H^2\sqrt{\omega_0}}t^{-(2+n)}}{\Omega_m+\Omega_r+\frac{1}{3}\left(\left(\frac{4\pi G}{3H^2}\right)n^2\Phi_0^2t_0^{2n}t^{-2(1+n)}-\frac{2n\Phi_0}{H^2}\left(\frac{pt_0^n}{\sqrt{\omega_0}}\right)t^{-(2+n)}+\frac{8\pi G}{3H^2}U_0t_0^{2(1+n)}t^{-2(1+n)}\right)}\right],
\end{equation}
which tends asymptotically to $-\frac{1}{2}$. Evaluating in the present time $t_0$ the expression \eqref{qcu12} leads to
\begin{equation}
    \label{qcu13}
    q_0=-\left[1-\frac{1}{2}\frac{\Omega_{m_{0}}+\frac{4}{3}\Omega_{r_0}+\left(1+\frac{1}{2\pi G\omega_0}\right)\frac{8\pi G}{3(H_0t_0)^2}n^2\Phi_0^2-(n(H_0t_0)+n(n+1))\frac{2\Phi_0}{3(H_0t_0)^2\sqrt{\omega_0}}}{\Omega_{m_0}+\Omega_{r_0}+\frac{1}{3}\left(\frac{4\pi G}{3(H_0t_0)^2}n^2\Phi_0^2-\frac{2n\Phi_0}{(H_0t_0)^2\sqrt{\omega_0}}+\frac{8\pi G}{3H_0^2}U_0\right)}\right],
\end{equation}
where $\Omega_{m_0}$ and $\Omega_{r_0}$ are the present density parameters for matter and radiation respectively. \\

Now, considering a slow roll of the field  $\phi$, the condition \eqref{qcu5} reduces to
\begin{equation}\label{qcu14}
    -\frac{8\pi G}{3}V(\phi)-2\ddot{\phi}-2H\dot{\phi}+4\dot{\phi}^2<0.
\end{equation}
In terms of the time, the expression \eqref{qcu14} reads
\begin{equation}\label{qcu15}
    -\left(\frac{8\pi G}{3}U_0t_0^2-4n^2\phi_0^2\right)t_0^{2n}t^{-2(1+n)}-\left(n(n+1)-2p\right)\phi_0t_0^nt^{-(2+n)}<0.
\end{equation}
For $\phi_0>0$ it can be reduced to the system
\begin{eqnarray}
 \label{qcu16}
 \frac{8\pi G}{3}U_0t_0^2-4n^2\phi_0^2>0,\\
 \label{qcu17}
 n(n+1)-2p>0.
\end{eqnarray}
Thus, it follows from \eqref{qcu16} that 
\begin{equation}\label{qcu18}
    U_0>\frac{3n^2\phi_0^2}{2\pi G t_0^2},
\end{equation}
and \eqref{qcu17} has for solutions $n<-2$ and $n>1$ for $p>1$, the latest corresponding to an accelerating expansion. The present value for the equation of state parameter for dark energy $\omega_{de}$ according to \eqref{qcu2}, \eqref{qcu8}, \eqref{qcu9}, \eqref{qcu10} and \eqref{qcu11}  reads
\begin{equation}
    \label{qcu19}
    \omega_{de_{0}}=-\frac{8\pi GU_0+n(n+1)(H_0t_0)^{-2}H_0^2\phi_0-4(H_0t_0)^{-1}H_0^2\phi_0-4n^2(H_0t_0)^{-2}H_0^2\phi_0^2}{8\pi GU_0-4(H_0t_0)^{-2}H_0^2\phi_0}.
\end{equation}
The equation \eqref{qcu19} can be written in the form
\begin{equation}
    \label{qcu20}
    4n^2(H_0t_0)^{-2}H_0^2\phi_0^2+(H_0t_0)^{-2}H_0^2[4(1+\omega_{de_{0}})-n(n+1)]\phi_0-8\pi G(1+\omega_{de_{0}})=0.
\end{equation}
Now, according to observational data Planck+SNe+BAO:  $\omega_{de_{0}}=-0.95\pm 0.080$ and from TT, TE, EE +lowE+lensing: $\Omega_{m_{o}}=0.3111\pm 0.0056$, $\Omega_{de_{0}}=0.6889\pm0.0056$, $\Omega_{r_{0}}=2.47\cdot10^{-5}\,h^{-2}$,  $H_{0}=67.4\pm0.5\,\frac{Km}{seg}Mpc^{-1}$ \cite{c26}. Thus from \eqref{qcu20} and  employing $H_0t_0\simeq \frac{2}{3}\Omega_{m_0}=0.2046^{+0.0074}_{-0.0066}$ \cite{c27} and taking the intermediate values for the observational parameters we arrive to the equation
\begin{equation}\label{qcu21}
  \phi_0=  \frac{2.016\cdot10^{-9}}{n^2}\left(-1.066\cdot10^7+6.2011\cdot10^7n^2+6.2011\cdot10^{7}n+\sqrt{\Delta_1(n)}\,\right),
\end{equation}
where 
\begin{equation}
    \label{qcu22}
    \Delta_1(n)=1.1376\cdot10^{14}+2.5225\cdot10^{15}n^2-1.3228\cdot10^{15}n+3.8453\cdot10^{15}n^4+7.6908\cdot10^{15}n^3+6.12719\cdot10^{11}n^2U_0.
\end{equation}
On the other hand, it follows from \eqref{qna10} that 
\begin{equation}\label{qcu23}
\Omega_{de_{0}}=\frac{(8\pi G)^2}{3H_0^2}U_0-\frac{32\pi G}{3(H_0t_0)}\phi_0.
\end{equation}
Solving \eqref{qcu23} por $U_0$ we obtain
\begin{equation}\label{qcu24}
    U_0=\frac{3H_0^2}{(8\pi G)^2}\left(\Omega_{de_{0}}+\frac{32\pi G}{3(H_0t_0)}\phi_0\right).
\end{equation}
Inserting the observational data the equation \eqref{qcu24} yields
\begin{equation}
    \label{qcu25}
    U_0=14.8633+3535.7293\,\phi_0.
\end{equation}
With the help of \eqref{qcu22} and \eqref{qcu25} the value of $\phi_0$ in order to achieve the observational value for $\omega_{de_{0}}$ is
\begin{equation}
    \label{qcu26}
    \phi_0=\frac{1}{n^2}\left(-0.01709+0.12501n^2+0.12501n+6.4512\cdot10^{-20}\sqrt{\Delta_2(n)}\,\right),
\end{equation}
where
\begin{equation}\label{qcu27}
    \Delta_2(n)=7.0305\cdot10^{34}+2.7366\cdot{36}n^2-1.02746\cdot10^{36}n+3.75517\cdot10^{36}n^4+7.51054\cdot10^{36}n^3.
\end{equation}
Finally, it follows from \eqref{qcu13} that the $\omega_0$ parameter of the model is given by
\begin{eqnarray}
   && \omega_0  = -\frac{3(H_0t_0)^2}{8\pi G}\frac{1}{n^2\phi_0^2}\frac{1}{q_0}\left[2(1+q_0)\left(\Omega_{m_0}+\Omega_{r_0}-\frac{2n\phi_0}{3(H_0t_0)^2}+\frac{8\pi G}{9H_0^2}U_0\right)-\left(\Omega_{m_0}+\frac{4}{3}\Omega_{r_0}\right)\right.\nonumber \\
    && \left.
    -\frac{4n^2\phi_0^2}{3(H_0t_0)^2}+n(n+1+H_0t_0)\frac{2\phi_0}{3(H_0t_0)^2}\right]. \label{qcu28}
\end{eqnarray}
Taking into account that according to Planck 2018 observational data $q_0=-0.5581^{+0.0273}_{-0.0267}$ the expression \eqref{qcu28} yields
\begin{eqnarray}
    \omega_0 &=& 1\cdot10^5\left[-5.016\cdot10^{13}n^2+2.1933\cdot10^{14}n^3+1.46\cdot10^{14}n-1.5832\cdot10^{9}n\sqrt{\Delta_3(n)}-1.7669\cdot10^{13}\,\right.\nonumber
    \\
    &&\left. +1.6685\cdot10^{9}\sqrt{\Delta_3(n)}+3.4194\cdot10^{14}n^4+4.4156\cdot10^{9}n^2\sqrt{\Delta_3(n)}\,
    \right]\left[7.7438\cdot10^{9}n^2+7.7438\cdot10^{9}n\right.\nonumber\\
    &&\left.-1.0586\cdot10^{9}+1\cdot10^{5}\sqrt{\Delta_3(n)}\,\right]^{-2},\label{qcu29}
\end{eqnarray}
where 
\begin{equation}
    \label{qcu30}
    \Delta_3(n)=5.9965\cdot10^{9}n^4+1.1993\cdot10^{10}n^3+4.37\cdot10^{9}n^2-1.6407\cdot10^{9}n+1.1226\cdot10^{8}.
\end{equation}
For example, when $n=3$, we obtain $\phi_0=0.3295\,M_p$ and $\omega_0=0.19925$. With these values our model is compatible with observational data. 

\section{Conclusion}

In this paper, in the framework of a recently introduced geometrical scalar-tensor theory of gravity, we propose a cosmological model that exhibits solutions describing the present period of accelerating expansion of the universe. In traditional scalar-tensor theories of gravity the background geometry is assumed apriori to be  Riemannian. This fact leads to the physical equivalence frame problem (the Jordan and Einstein frames) \cite{c21}. In the geometrical gauge scalar-tensor theory the equivalence frame problem is avoided  as a consequence of determining  the background geometry of the theory by imposing the Palatini's variational principle, instead of regarding the background geometry as Riemannian,   apriori. Thus for a traditional scalar-tensor theory of gravity the background geometry resulting  is of  the Weyl-Integrable type, whose symmetry group is the Weyl group of transformations. The problem is that the original scalar-tensor action is not an invariant under the new geometrical symmetry group. Hence, to solve the problem it is proposed a new scalar-tensor action that generates a new kind of scalar-tensor theory called geometrical gauge scalar-tensor theory of gravity.  \\

Working with an invariant metric under the Weyl group, we have derived a cosmological model for the present epoch of accelerating expansion in which the dark energy sector is described by the Weyl scalar field, which is geometrical in origin. The deceleration parameter resulting tends asymptotically to $-1/2$. We study the case of a power law expanding universe and we obtain a power law solution for the Weyl scalar field. To illustrate how our formalism works we have study the simplest case for a geometrical scalar-tensor theory which is when $\omega(\phi)=\omega_0=const$.\\

We have obtained a potential for the Weyl scalar field resulting of the power law type. The deceleration parameter and the equation of state parameter for the present time are in agreement with observational data according to the PLANCK 2018 results for specific values of $\phi_0$ and $\omega_0$, which for a power $n=3$ of the scalar field we obtain $\phi_0=0.3295\,M_p$ and $\omega_0=0.19925$.\\

Therefore, an accelerating expanding cosmological scenario can be well-described by a geometrical scalar-tensor theory of gravity, where the Weyl scalar field has a geometrical origin in the sense that it originally  forms part of the affine structure of the space-time manifold. Notice that only for the class of observers that use the invariant metric the Weyl scalar field becomes physical, because for these observers the background geometry can be seen as Riemannian effective. 

\section*{Acknowledgements}

\noindent J. E. Madriz-Aguilar, M. Montes and A. Bernal acknowledge  CONACYT
M\'exico and Centro Universitario de Ciencias Exactas e Ingenierias of Guadalajara University for financial support.


\bigskip

\begin{thebibliography}{99}
	
\bibitem{c1} A. G. Riess et al., Astron. J {\bf 116} (1998) 1009.
\bibitem{c2}  S. Perlmutter et al., Nature {\bf 391} (1998) 51.
\bibitem{c3}  D. N. Spergel et al., Astrophysics J. Supp. {\bf 170} (2007)
377.
\bibitem{c4}  M. Betoule et al., Astron. Astrophys. {\bf 568} (2014) A22.
SDSS collaboration.
\bibitem{c5} J. Frieman, M. Turner, D. Huterer, Ann. Rev. Astron. Astrophys. {\bf 46} (2008) 385-432.
\bibitem{c6} J. Martin, Comptes Randus Physique 13 (2012) 566-665.
\bibitem{c7} J. S. Peracaula, "The Cosmological Constant Problem and Running Vacuum in the Expanding Universe" (2022). 	ArXiv:2203.13757 [gr-qc].
\bibitem{c8} J. E. Madriz Aguilar, M. Bellini, M. A. S. Cruz, 	Grav. Cosmol. {\bf 14} (2008)286-291.
\bibitem{c9} J. E. Madriz-Aguilar,  J. Zamarripa, A. Peraza, J. A. Licea, Phys. Dark. Univ. {\bf 18} (2017) 11-16.

\bibitem{c10} J. S. Alcaniz, Braz. J. Phys. {\bf 36} (2006) 1109.

\bibitem{c11} J. E. Madriz-Aguilar, J. Zamarripa, M. Montes, J. A. Licea, C. de Loza, A. Peraza, The European Phys. J. Plus {\bf 137} (2022) 1, 135.
\bibitem{c12}  R. de Putter and E. V. Linder, Astropart. Phys. {\bf 28} (2007) 263-72.

\bibitem{c13} J. E. Madriz-Aguilar, M. Montes, Phys. Dark Univ. {\bf 21} (2018) 47-54.

\bibitem{c14} L. F. P. da Silva, J. E. Madriz Aguilar, Mod. Phys. Lett. A {\bf 23} (2008) 1213-1221.

\bibitem{c15} T. S. Almeida, M. L. Pucheu, C. Romero, J. B. Formiga, Phys. Rev. D {\bf 89} (2014) 064047.

\bibitem{c16} M. L. Pucheu, F. A. P. Alves Junior, A. B. Barreto, C. Romero, Phys. Rev. D {\bf 94} (2016) 064010.

\bibitem{c14} J. E. Madriz Aguilar, A. Bernal, M. Montes, J. Zamarripa, E. Aceves, Phys. Dark Univ. {\bf 35} (2022) 100988.

\bibitem{c15} M. Montes, Jos\'e Edgar Madriz Aguilar, and V. Granados, Can. J. Phys. \textbf{97} (2019) 517-523.

\bibitem{c16} J. E. Madriz-Aguilar, J. Zamarripa, M. Montes, C. Romero, Phys. Dark Univ. 28 (2020) 100480.

\bibitem{c17} 
J. E. Madriz Aguilar, C. Romero, J. B. Fonseca Neto, T. S. Almeida, J. B. Formiga, Class. Quant. Grav. {\bf 32} n°21 (2015) 215003.

\bibitem{c18} J. E. Madriz Aguilar, A. G. Ocaranza, M. Montes, J. Zamarripa, Phys. Dark Univ. {\bf 30} (2020) 100706.

\bibitem{c19} Y. Liu, Eur. Phys. J. Plus (2021) 136: 901.

\bibitem{c20} J. L. Rosa, J. P. S. Lemos, Phys. Rev. D 104, 124076 (2021).

\bibitem{c21} I. Quiros, R. García-Salcedo, J. E. Madirz-Aguilar, T. Matos, Gen. Relat.  Grav. {\bf 45} n° 2 (2013) Page 489-518.

\bibitem{c22} M. Bellini, J. E. Madriz-Aguilar, M. Montes, p. A. Sánchez, Phys. Dark Univ. {\bf 25} (2019) 100309.

\bibitem{c23} L. S. Ridao, M. Bellini, Phys. Lett. B {\bf 751} (2015) 565-571.

\bibitem{c24} A. Delhom, I. P. Lobo, G. J. Olmo, C. Romero, Eur. Phys. J. C {\bf 79} (2019) 878.

\bibitem{c25} E. Scholz, "A Weyl geometric scalar field approach to dark sector" (2022). ArXiv:2202.13467 [gr-qc].

\bibitem{c26}  N. Aghanim et al. (Planck Collaboration), (2018) ArXiv:
1807.06209.  

\bibitem{c27}  Particle Data Group, Phys. Rev. D98 (2018) 030001.


\end{thebibliography}
\end{document}